\documentclass[11pt,fleqn]{article} 
\usepackage{graphicx} 
\usepackage{amsfonts} 
\usepackage{amssymb} 
\usepackage{epsfig}
\usepackage{cite}
\usepackage{user5}  

\newcommand{\rsm}{\ensuremath{{\rm SM}}}
\newcommand{\rmssm}{\ensuremath{{\rm MSSM}}}
\newcommand{\rsusy}{\ensuremath{{\rm SUSY}}}

\newcommand{\tbeta}{\ensuremath{\tan\beta}}
\newcommand{\mSM}{\ensuremath{M_{H_{SM}}}}

\newcommand{\ma}{\ensuremath{M_{A^0}}}

\newcommand{\Mh}{\ensuremath{M_{h^0}}}
\newcommand{\mz}{\ensuremath{M_{Z}}}
\newcommand{\mw}{\ensuremath{M_{W}}}
\newcommand{\mh}{m_{h^0}}

\newcommand{\msto}{m_{\tilde{t}_1}}
\newcommand{\mstt}{m_{\tilde{t}_2}}

\def\cw{c_{{\scriptscriptstyle W}}}
\newcommand{\mhtree}{M^{2 \,{\mathrm{tree}}}_{h^0}}
\newcommand{\be}{\begin{equation}}
\newcommand{\ee}{\end{equation}}
\newcommand{\bea}{\begin{eqnarray}}
\newcommand{\eea}{\end{eqnarray}}

\begin{document} 
\hfill{} 
\begin{center}
\textbf{\large Quantum effects to the Higgs boson self-couplings in the 
$\rsm$ and in the $\rmssm$}\vspace*{0.4cm}\\
{\par\centering 
A.~Dobado$^{\,\small{a}}$\,, M.J.~Herrero $^{\,\small{b}}$\,,
W.~Hollik $^{\,\small{c}}$ and 
{\underline{S.~Pe{\~n}aranda}} $^{\,\small{c,\,d}}$
\vspace*{0.2cm}~\footnote{Talk given by S.P. at SUSY02, The 10th International
Conference on Supersymmetry and Unification of Fundamental Interactions, DESY Hamburg, Germany.\\
\hspace*{0.6cm}The work of S.P. has been partially 
supported by the \textit{Fundaci{\'o}n Ram{\'o}n Areces}.}
~\footnote{e-mails:
malcon@fis.ucm.es, herrero@delta.ft.uam.es,
hollik@mppmu.mpg.de, siannah@mppmu.mpg.de}\\
\par} 
{\par\centering 
$^{\small{a}}$
\textit{Departamento de F{\'\i}sica Te{\'o}rica,\\
Universidad Complutense de Madrid, 28040 Madrid, Spain \vspace*{0.1cm}}}
{\par\centering 
$^{\small{b}}$
\textit{Departamento de F{\'\i}sica Te{\'o}rica,\\
Universidad Aut{\'o}noma de Madrid, Cantoblanco, 28049 Madrid, 
Spain \vspace*{0.1cm}}}
{\par\centering 
$^{\small{c}}$
\textit{Max-Planck-Institut f\"ur Physik,\\
F\"ohringer Ring 6, D-80805 M\"unchen, 
Germany\vspace*{0.1cm}}}
{\par\centering 
$^{\small{d}}$
\textit{Institut f\"{u}r Theoretische Physik,\\ Universit\"{a}t Karlsruhe,
Kaiserstra\ss{}e 12, D--76128, Germany}}
\end{center}
\vspace*{0.3cm}
{\par\centering\textbf{\large Abstract}\\ 
\par} 
\noindent We show that the effects of 
heavy Higgs particles and heavy top-squarks 
in the one-loop self-couplings of
the lightest CP-even $\rmssm$ Higgs boson 
decouple from the low energy theory 
when the self-couplings are
expressed in terms of the Higgs boson mass $\Mh$.
Our conclusion is that the $h^{0}$ self-interactions become very close 
to those of the $\rsm$ Higgs boson
and, therefore, MSSM quantum effects could
only be revealed by very high precision experiments. 
\vspace*{0.1cm}\\
\noindent KA-TP-12-2002,  MPI-PhT/2002-58, FTUAM/02-27, IFT/CSIC-UAM-02-45,
hep-ph/0210315


\renewcommand{\thefootnote}{\arabic{footnote}}
\setcounter{footnote}{0}

\section{Introduction} 

One of the most important issues at the next generation of colliders is the 
discovery of at least one light Higgs boson particle and the elucidation of
the mechanism of symmetry breaking~\cite{WGHiggs}. 
Particularly relevant in order to establish the Higgs mechanism 
experimentally in an unambiguous way, is the reconstruction of the Higgs
self-interaction potential. This task requires the
measurement of the trilinear and quartic self-couplings as predicted in the
Standard Model ($\rsm$) or in supersymmetric (SUSY) theories.
Since the predictions of these self-couplings are different in both theories, 
their experimental measurement could provide not just an essential way
to determine the mechanism for generating the masses of the fundamental
particles but also an indirect way to test SUSY. 
The cross section for double
Higgs production (e.g., $Zhh$) is related to the triple Higgs self-couplings 
$\lambda_{hhh}$, which in turn is related to the spontaneous symmetry breaking
shape of the Higgs potential. Experimental studies indicate that for a 
$\rsm$-like Higgs boson with $m_h=120$ GeV at $1000 fb^{-1}$, a
precision of $\delta \lambda_{hhh}/\lambda_{hhh}= 23\%$
is possible at TESLA~\cite{exp}. Strategies for
measuring the $\rsm$ Higgs boson self-couplings at the
LHC have been also discussed recently in~\cite{rainwater}.
Many other studies have addressed the issue of the
measurement of the neutral 
Higgs self-couplings in the Minimal Supersymmetric Standard Model ($\rmssm$) and 
also in the two-Higgs doublet model (2HDM)~\cite{regions,RadCorCouplings}. 

In recent papers, we investigated how far the $\rmssm$ Higgs
potential reproduces the $\rsm$ potential when the non-standard
particles are heavy~\cite{Selfmt4,MAHyo}. More concretely, we explore the
decoupling behaviour of the radiative
corrections to the $h^{0}$ self-couplings at the one-loop level, 
both numerically and analytically. A first step into
this direction was the analysis of the leading Yukawa contributions of order 
${\cal O}(m_t^4)$ to the lightest $\rmssm$ Higgs boson
self-couplings~\cite{Selfmt4}. Recently, an analytical study of the
contributions from the $\rmssm$ Higgs sector itself~\cite{MAHyo} 
has been done. This talk gives a summary on the results of these calculations.

First, we present in Table 1 the trilinear and quartic $\rsm$ and $\rmssm$ 
Higgs boson self-couplings, at the tree-level.
Clearly, for arbitrary values of the $\rmssm$ Higgs-sector input parameters, 
$\tan\beta$ and $\ma$, the values of the $h^{0}$ self-couplings are different 
from those of the $\rsm$ Higgs boson. However, the
situation changes in the so-called {\it decoupling limit} of the Higgs sector
where $\ma \gg \mz$, yielding a particular spectrum with heavy
$H^0\,,H^\pm\,,A^0$ Higgs bosons with degenerate masses, and a light $h^0$ boson having a
tree-level mass of $\mh \backsimeq \mz |\cos 2 \beta|$~\cite{dec}. In this limit, which
also implies $\alpha\rightarrow \beta-\pi/2$, the $h^0$ self-couplings converge
respectively to
$\lambda_{hhh}^0 \simeq\,3 g/2\,\mw \,\mhtree\,$ and 
$\lambda_{hhhh}^0 \simeq \,3\, g^2/4\,\mw^2 \mhtree$ and therefore, they
converge to their respective $\rsm$ Higgs boson self-couplings with
the same mass. Thus, we can conclude that, at the tree level, there is
decoupling of the heavy $\rmssm$ Higgs sector in the $\ma \gg \mz$
limit. Therefore, by
studying the light Higgs boson self-interactions it will be very difficult to
unravel its SUSY origin.
This is another reason to investigate the effects of the quantum contributions
to these self-interactions.

{\small{
\begin{table}[thb]
\vspace*{-0.1cm}
\renewcommand{\arraystretch}{2.0}
\begin{center}
\begin{tabular}{|lc||cc|} \hline
  \multicolumn{2}{|c||}{\large{$\phi$}} &\large{$\lambda_{\phi \phi \phi}$} & 
 \large{$\lambda_{\phi \phi \phi \phi}$} \\
\hline \hline
\large{$\rsm$}~ &\large{$H$} & \large{$\frac{3g M_H^2}{2\,\mw}$}&
\large{$\frac{3g^2 M_H^2}{4\,\mw^2}$} \\ \hline
\large{$\rmssm$}~ & \large{$h^o$} & 
\large{$ \,\frac{3 g \mz}{2 \cw}$}
\large{$\cos2\alpha \sin(\beta + \alpha)$} &
\large{$ \,\frac{3 g^2}{4 \cw^2}$}\large{$\cos^2 2\alpha$} \\ \hline
\end{tabular}
\renewcommand{\arraystretch}{1.2}\vspace*{0.1cm}\\
Table 1: Tree-level $\rsm$ and $\rmssm$ Higgs boson self-couplings\vspace*{-1.0cm}
\end{center}
\end{table}}}

\section{One-loop Higgs sector contributions in the {\it decoupling limit}}
\label{sec:GreenMSSM}

We summarize here the one-loop results for the $h^0$ self-couplings from the
$\rmssm$ Higgs sector itself by considering the already described {\it decoupling limit}~\cite{MAHyo}.
We only consider
the set of diagrams that provides contributions to the $\rmssm$ $h^0$
self-couplings different from the $\rsm$ ones.
We have checked explicitly that the one-loop contributions from diagrams that have
at least one gauge boson flowing in the loops are the same in both models. 
Consequently, we are considering, first,
diagrams involving only the $\rmssm$ heavy Higgs bosons ($H^0\,,H^\pm\,,$ and
$A^0$), second, diagrams with these heavy particles and the
Goldstone bosons or the lightest Higgs boson appearing in the same loop, 
and finally, 
diagrams involving just Goldstone bosons and the lightest Higgs boson
in the loops ({\it light} contributions); all of them are,
in principle, different in the two  models. 
However, as explained in detail in~\cite{MAHyo}, 
we found that the one-loop {\it light} contributions approach
the $\rsm$ ones in the $\ma \gg \mz$ limit and therefore, they 
become indistinguishable in both models. 
This is equivalent to saying that the
difference between the one-loop unrenormalized vertex functions 
of the
two theories in the {\it decoupling limit} is coming from
diagrams including at least one heavy $\rmssm$ Higgs particle. 

The results for the one-loop contributions to the $n$-point vertex functions,
$\Delta \Gamma_{h^{0}}^{\,(n)}$, 
coming from the Higgs sector itself, 
can be summarized generically as
\begin{eqnarray}
\label{eq:oneloop}
\Delta \Gamma_{h^{0}}^{(n)}&\simeq&\mz^2\,\left[{{\cal{O}}\left({\frac{1}{\epsilon}}\right)+
{\cal{O}}\left(\log\frac{M_{EW}^2}{\mu_{0}^2}\right)+
{\cal{O}}\left(\log\frac{\ma^2}{\mu_{0}^2}\right)+
\normalsize{{\mbox{ finite terms }}}}\right]\nonumber\\
&&+\ma^2\,\left[{{\cal{O}}\left({\frac{1}{\epsilon}}\right)+
{\cal{O}}\left(\log\frac{\ma^2}{\mu_{0}^2}\right)+
\normalsize{{\mbox{ finite terms }}}}\right]\,,
\end{eqnarray}
where $M_{EW}^2 \equiv \mz^2\,,\mw^2\,,\mh^2$.
These contributions are UV-divergent and contain both a  
logarithmic dependence and a 
quadratic dependence on the heavy pseudoscalar mass $\ma$. 
Therefore, all potential
non-decoupling effects of the heavy $\rmssm$ Higgs particles  manifest themselves 
as divergent contributions in $D=4$ and some finite contributions,
logarithmically and quadratically dependent on $\ma$.

As a next step, renormalization has to be done
to get finite vertex functions and physical observables; 
this is performed in practice by adding appropriate counterterms.
The results of the vertex counterterms
in the {\it decoupling limit} as well as the results for the renormalization constants,
obtained by using the on-shell scheme~\cite{Renor}, 
have been presented in~\cite{MAHyo}. 
The results for the renormalized
vertex functions, $\Delta {\Gamma_{\,R\,\,h^{0}}^{\,(n)}}$, 
defined by summing $\Delta \Gamma_{h^{0}}^{\,(n)}$
and the counterterm contributions, can be expressed as follows,
\bea
\label{eq:firstredef}
\Delta {\Gamma_{\,R\,\,h^{0}}^{\,(2)}}&=&\Delta \Mh^2\,,\nonumber\\
\Delta {\Gamma_{\,R\,\,h^{0}}^{\,(3)}}&=&\frac{3g}{2\mz \cw}\, \Delta
\Mh^2+\frac{g^3}{64 \pi^2 \cw^3}\,\mz\,\cos^2 2\beta\,\Psi_{\rmssm}^{\rm {rem}}\,,\nonumber\\
\Delta {\Gamma_{\,R\,\,h^{0}}^{\,(4)}}&=&\frac{3g^2}{4\mz^2\cw^2}\,\Delta \Mh^2+
\frac{g^4}{64\pi^2\cw^4 }\,\cos^2 2\beta\,\Psi_{\rmssm}^{\rm {rem}}\,,
\eea
where $\Delta\Mh^2$ represents the $h^{0}$ mass-squared correction 
for the $h^{0}$, 
\begin{equation}
\label{eq:masaandremain}
\Delta\Mh^2=\mz^2\,\left[{\cal{O}}\left(\frac{1}{\epsilon}\right)+
{\cal{O}}\left(\log\frac{M_{EW}^2}{\mu_{0}^2}\right)+
{\cal{O}}\left(\log\frac{\ma^2}{\mu_{0}^2}\right)+
\normalsize{{\mbox{ finite terms }}}\right]\,,
\end{equation}
and  $\Psi_{\rmssm}^{\rm {rem}}$ refers to the remaining finite terms resulting 
exclusively
from the {\it light} contributions. 
For the discussion, we have dropped those terms which are identical in the SM and the MSSM,
i.e.\ the pure gauge part including the pure Goldstone contributions.
As a consequence of considering not a complete set of one-loop diagrams, 
the mass correction $\Delta\Mh^2$ is UV-divergent.
In order to cancel this residual divergence in the renormalized 
two-point function, it is necessary to include also the
diagrams dropped here. 

The mass correction~(\ref{eq:masaandremain}) contains
finite terms proportional to $\mz^2$ and 
logarithmic dependences on the heavy mass $\ma$
as well as on the electroweak masses. 
The quadratic heavy-mass terms $\sim\ma^2$, however, appearing  in the
unrenormalized vertex functions~(\ref{eq:oneloop}), disappear in
the on-shell renormalization procedure. 
Once the tree-level Higgs-boson mass is replaced by the corresponding 
one-loop mass, $\Mh^2 ={\Mh^2}^{\rm tree}+\Delta \Mh^2\,,$
with $\Delta \Mh^2$ given in~(\ref{eq:masaandremain}), we obtain that  
the singular ${\cal O}(1/\epsilon)$ terms and the logarithmic heavy-mass
terms also disappear in the renorma\-lized 3- and 4-point functions. 
Hence, we have an analytic demonstration 
that the heavy $\rmssm$ Higgs bosons decouple in the $\ma \gg \mz$ limit.

Nevertheless, after the previously commented terms in the
radiative corrections to the trilinear and quartic $h^{0}$ self-couplings have
been absorbed in the $h^{0}$ mass redefinition, other finite terms, contained in
$\Psi_{\rmssm}^{\rm {rem}}$  in~(\ref{eq:firstredef}), still do remain
and  could give rise to differences between the predictions of the $\rmssm$ and the $\rsm$.
For the interpretation of these remaining terms
it is thus crucial to perform the corresponding one-loop analysis 
for the  self-interactions of the Higgs boson in the SM as well.
After renormalization of the trilinear and quartic self-couplings in the $\rsm$, 
done in \cite{MAHyo},
we find
\be
\label{eq:firstredefSM}
\Delta {\Gamma_{\,R\,\,H_{SM}}^{\,(3)}}=
\frac{g^3}{64 \pi^2 \cw^3}\,\mz\,\Psi_{SM}^{\rm{rem}}\,,\,\,\,\,\,\,\,\,
\Delta {\Gamma_{\,R\,\,H_{SM}}^{\,(4)}}=
\frac{g^4}{64 \pi^2 \cw^4}\,\Psi_{SM}^{\rm{rem}}\,\,,
\ee
with the $M_{H_{SM}}$-dependent function $\Psi_{SM}^{\rm{rem}}$
representing 
the only remaining finite terms in the renormalized $H_{\rsm}$ self-couplings. 
In general, these finite contributions are different from the finite 
$\Psi_{\rmssm}^{\rm{rem}}$ terms originating from the 
{\it light} contributions in the $\rmssm$. However, by identifying 
${M^{{\mathrm{tree}}}_{h^0}}^2 \simeq \mz^2 \cos^2 2\beta 
\longleftrightarrow \mSM^2$
in the {\it decoupling limit}, we obtain the asymptotic relation
$\Psi_{MSSM}^{\rm{rem}}\longrightarrow \Psi_{SM}^{\rm{rem}}$.
Therefore, these EW-finite terms are common to both
the lightest $\rmssm$ Higgs boson, $h^{0}$, and the $\rsm$  $H_{\rsm}$, 
in the case $\ma \gg \mz$.

Thus, we have shown that the $\rmssm$ {\it heavy} Higgs one-loop
contributions can be absorbed in the redefinition of the lightest Higgs boson mass $\Mh$
and therefore, decoupling of the heavy Higgs particles occurs.
Similarly, the divergent part of the {\it light} contributions as well as part of the
finite terms are also absorbed in this $\Mh$ mass correction. Another part of the
finite terms in the renormalized 3- and 4-point $\rmssm$ 
vertex functions~(\ref{eq:firstredef}) remains. 
These remaining contributions, however, coincide 
with the corresponding $\rsm$ ones in the $\ma \gg \mz$ limit by identifying $\mSM$ with $\Mh$ and, 
therefore, 
they drop out when differences in the  predictions of both models are considered.
Consequently, 
the trilinear and quartic $h^0$ self-couplings at the
one-loop level and in the $\ma \gg \mz$ limit 
have the same structure as the SM self-couplings~\cite{MAHyo}. 

\section{${\cal{O}} (m_t^4)$ one-loop contributions}
\label{sec:selfcouplings}

The one-loop leading Yukawa corrections from top and stop loop contributions to the renormalized
$h^0$ vertex functions were derived in~\cite{Selfmt4} by considering the 
{\it{decoupling limit}} and a heavy top-squark sector, with $\tilde{t}$ masses
large as compared to the electroweak scale.
Two different scenarios have been addressed: 
First, in an analytical study, the two stop masses 
are heavy but close to each other, i.e. 
$|\msto^2-\mstt^2| \ll |\msto^2+\mstt^2|\,$~\cite{TesisS}. Second,
the other possible scenario where the stop mass splitting is of the order of 
the $\rsusy$ mass scale, which corresponds to
$ |\msto^2-\mstt^2| \simeq |\msto^2+\mstt^2|\,\,,$ in a numerical analysis.

In the analytical studies, we found that the one-loop contribution to the
renormalized two-point function is given by
$\Delta {\Gamma}_{R\, h^{0}}^{\,t,\tilde{t}\,(2)}=\Delta \Mh^2$, with
$\Delta \Mh^2$ being the (leading) one-loop correction to the $h^0$ mass-squared,  
\be
\label{eq:deltamh}
\Delta \Mh^2=-\frac{3}{8 \pi^2}\frac{g^2}{\mw^2}\,m_t^4\,
\log \frac{m_t^2}{m_{\tilde t_1} m_{\tilde t_2}} \,.
\ee
Thus, we were able to write the results for the trilinear and quartic 
self-couplings, obtained as the corresponding renormalized vertex
functions, in the following way, 
\bea
\label{eq:redefinition34p}
&& \Delta {\Gamma}_{R\, h^{0}}^{\,t,\tilde{t}\,\,(3)} =
 \frac{3}{v}\,\Delta \Mh^2 
-\frac{3}{8 \pi^2}\frac{g^3}{\mw^3}\,m_t^4 \, , \nonumber \\
&& \Delta {\Gamma}_{R\, h^{0}}^{\,t,\tilde{t}\,\,(4)} =
\frac{3}{v^2}\,\Delta \Mh^2 
-\frac{3}{4 \pi^2}\frac{g^4}{\mw^4}\,m_t^4 \,.
\eea

The UV-divergences are canceled by the renormalization procedure, and the 
logarithmic heavy mass term, which looks like a non-decoupling effect 
of the heavy particles in the renormalized vertices, disappears when the 
vertices are expressed in terms of the
Higgs-boson mass shift~(\ref{eq:deltamh}). Therefore, 
they do not appear directly in related observables, i.e.\ they decouple. 
However, some non-logarithmic terms ${\cal{O}} (m_t^4)$ remain in
the trilinear and quartic $h^{0}$ self-couplings~(\ref{eq:redefinition34p}).
Without these top-mass terms, the self-couplings at the
one-loop level have the same form as the tree-level couplings,
with the tree-level Higgs-boson mass replaced by the corresponding
one-loop mass, $\Mh^2 = \mhtree + \Delta\Mh^2 \,.$

By deriving the equivalent one-loop ${\cal{O}} (m_t^4)$ contributions in the $\rsm$ we found that,
after on-shell renormalization of the trilinear and quartic
couplings in the $\rsm$, the results correspond precisely to the two non-logarithmic terms 
in~(\ref{eq:redefinition34p}). Hence, 
these top-mass  terms are common 
to both $h^0$ and $H_{SM}$. Therefore, we conclude that the ${\cal O} (m_t^4)$
one-loop contributions to the MSSM $h^{0}$ vertices {\it{either}}
represent a shift in the $h^{0}$ mass and 
in the $h^0$ triple and quartic self-couplings, which can be
absorbed in $\Mh$, {\it{or}} reproduce the $\rsm$ top-loop corrections. 
The triple and quartic $h^0$ couplings thereby 
acquire the structure of the $\rsm$ Higgs-boson self-couplings. 
These results have been confirmed also numerically in~\cite{Selfmt4}.

\begin{figure}[t]
\begin{center}
\begin{tabular}{cc}
\resizebox{7.5cm}{!}{\includegraphics{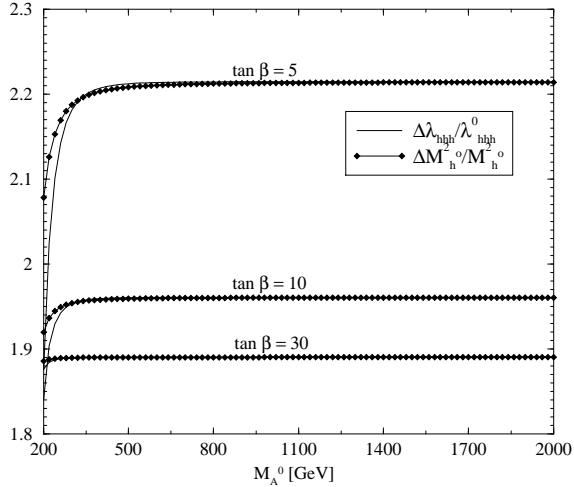}}
\end{tabular}
\end{center}\vspace*{-0.8cm}
\caption{${\cal O}(m_t^4)$ radiative corrections
to the trilinear $h^{0}$ self-coupling and to the $h^{0}$ mass as a
  function of  $\ma$, when the stop mass splitting is of ${\cal{O}}(M_{\tilde Q})$.\vspace*{-0.5cm}}
\label{fig:decouple}
\end{figure}

The other scenario where the stop-mass splitting is of the order of 
the SUSY mass scale, has been analyzed numerically,
based on the exact results for ${\cal O}(m_t^4)$ corrections to the triple and quartic self-couplings.
Details of this analysis can be found in~\cite{Selfmt4}. Here 
 we present in Fig.~\ref{fig:decouple} one example of the numerical results for the 
varia\-tion of the trilinear coupling and for the ${\cal O}(m_t^4)$
$h^{0}$ mass correction as functions of $\ma$, for different values of $\tbeta$, by choosing 
the set of SUSY input parameters to be 
$M_{\tilde Q}\sim 15 {\mbox{ TeV}}\,,\,
M_{\tilde U} \sim \mu \sim |A_t| \sim 1.5 {\mbox{ TeV}}\,$,
such that one gets $|\msto^2-\mstt^2|/|\msto^2+\mstt^2| \simeq 0.97$.
The radiative correction to the angle $\alpha$ is also taken into account and
the non-logarithmic finite contribution to the three-point function 
owing to the top-triangle diagrams is not taken into account in the figure since,
as we mentioned before, it
converges always to the $\rsm$ term. We see in this figure that 
for very large SUSY scales, 
the relation $\Delta \lambda_{hhh}/{\lambda_{hhh}^0}
\approx \Delta \Mh^2/{\mhtree}$ is again fulfilled.
Quantitatively, one finds
that for $\tbeta=5$ and $\ma=2$ TeV, the difference between vertex and mass corrections
is  $\sim 0.03\%$, and for
the most unfavo\-rable case, i.e $\tbeta=5$ and $\ma=200$ GeV, it is about $\sim 0.2\%$.

Therefore, from the numerical analysis one 
can conclude that also for the case of a heavy stop system with large
mass splitting, the ${\cal O}(m_t^4)$  corrections to the trilinear 
$h^{0}$ self-couplings are absorbed to the largest extent in the 
loop-induced shift of the $h^{0}$ mass, leaving only a very
small difference, which can be interpreted as 
the genuine one-loop correction
when $\lambda_{hhh}$ is expressed in terms of $\Mh$.
Similar results have been obtained also for the 
quartic $h^{0}$ self-coupling.

\section{Conclusions}
\label{sec:conclu}

\hspace*{0.5cm} We showed that
the one-loop Higgs-sector corrections and the ${\cal{O}}(m_t^4)$  Yukawa contributions 
to the lightest MSSM Higgs-boson self-couplings disappear to a large extent
when the self-couplings are expressed in terms of the $h^{0}$-boson mass, in the 
limit of large $\ma$ and heavy top squarks, leaving behind the quantum corrections
of the SM.
Therefore, the triple and quartic $h^{0}$ self-couplings acquire the structure of the $\rsm$ Higgs-boson 
self-couplings. 

Equivalently, we have demonstrated that  heavy 
Higgs particles and heavy top-squarks decouple from the low energy, 
at the electroweak scale and at one-loop level, and  the $\rsm$ Higgs sector 
is recovered also in the Higgs self-interactions. 
Consequently, we would need high-precision experiments for an 
experimental verification of the supersymmetric nature of the
Higgs-boson self-interactions. 

\begingroup\raggedright

\end{document}